\documentclass[3p,times,procedia]{elsarticle}
\usepackage{nupha_ecrc}
\usepackage{hyperref}
\volume{00}

\firstpage{1}

\journalname{Nuclear Physics A}

\runauth{J. Wilkinson for the ALICE Collaboration}

\jid{nupha}

\jnltitlelogo{Nuclear Physics A}

\usepackage{amssymb}
\biboptions{sort&compress}

\usepackage[figuresright]{rotating}

\usepackage{xspace}
\newcommand{\pp}{\mbox{pp}\xspace}

\newcommand{\PbPb}{\mbox{Pb--Pb}\xspace}
\newcommand{\pPb}{\mbox{p--Pb}\xspace}
\newcommand{\sqrts}{\sqrt{s}}
\newcommand{\sqrtsNN}{\ensuremath {\sqrt{s_{\rm NN}}}}

\newcommand{\TeV}{\ensuremath{\mathrm{TeV}}\xspace}

\newcommand{\GeVc}{\ensuremath{\mathrm{GeV}/c}\xspace}

\newcommand{\pt}{\ensuremath{p_{\rm T}}\xspace}
\newcommand{\pT}{\pt}

\newcommand{\etalab}{\eta}

\newcommand{\DtoKpi}{{\rm D}^0 \to {\rm K}^-\pi^+}
\newcommand{\DtoKpipi}{{\rm D}^+\to {\rm K}^-\pi^+\pi^+}
\newcommand{\DstartoDpi}{{\rm D}^{*+} \to {\rm D}^0 \pi^+}

\newcommand{\DstophipitoKKpi}{{\rm D_s^{+}\to \phi\pi^+\to K^-K^+\pi^+}}
\newcommand{\Dzero}{\ensuremath{\rm D^0}\xspace}

\newcommand{\Dstar}{\ensuremath{\rm D^{*+}}\xspace}

\newcommand{\Dplus}{\ensuremath{\rm D^+}\xspace}

\newcommand{\dEdx}{{\rm d}E/{\rm d}x}

\newcommand{\RpPb}{\ensuremath{R_\mathrm{pPb}}\xspace}
\newcommand{\RAA}{\ensuremath{R_\mathrm{AA}}\xspace}
\newcommand{\Raa}{\RAA}
\newcommand{\QpPb}{\ensuremath{Q_\mathrm{pPb}}\xspace}
\newcommand{\TpPb}{\ensuremath{T_\mathrm{pPb}}}

\newcommand{\figref}[1]{Fig.~\ref{#1}}

\newcommand{\Figref}[1]{Figure~\ref{#1}}

\begin{document}

\begin{frontmatter}

\dochead{Quark Matter 2015}

\title{Measurements of heavy-flavour production in \pPb~collisions with ALICE}

\author{Jeremy Wilkinson}
\ead{jeremy.wilkinson@cern.ch}
\author{for the ALICE Collaboration}
\address{Physikalisches Institut, Ruprecht-Karls-Universit\"{a}t Heidelberg, Germany}

\begin{abstract}

The production of open heavy-flavour particles was studied in \pPb~collisions at$\sqrtsNN=5.02$~TeV using the ALICE detector. Three separate observables were used: the hadronic decays of D mesons at mid-rapidity, and semileptonic decays of heavy-flavour hadrons to electrons and muons at mid-rapidity and forward rapidity, respectively. The most recent ALICE measurements of the nuclear modification factor, $\RpPb$, of open charm and beauty are reported, along with the centrality and multiplicity dependence of D-meson production in \pPb~collisions.
\end{abstract}


\end{frontmatter}

\section{Introduction}

\label{sec:Intro}
Measurements of heavy-quark (charm and beauty) production offer a unique probe of the properties of the Quark-Gluon Plasma (QGP) produced in ultrarelativistic heavy-ion collisions. 
Due to their large masses, charm and beauty quarks are predominantly produced in the initial hard scatterings of a heavy-ion collision, rather than in thermal processes at later times, meaning that they experience the full evolution of the collision system, and also that their production is described well by calculations of perturbative Quantum Chromodynamics (pQCD). 
Measurements of open heavy-flavour production in \pPb~collisions make it possible to disentangle Cold Nuclear Matter (CNM) effects, such as transverse momentum broadening and the modification of nuclear parton distribution functions, from final-state effects that occur in the Quark-Gluon Plasma (QGP) formed in \PbPb~collisions.

A brief overview of the detectors and analysis strategies used for heavy-flavour measurements is given in the following paragraphs; full details of the ALICE detector and its performance can be found in~\cite{ALICEperformance}.

Prompt D mesons~\cite{ALICEDmespPbmb} are reconstructed via the hadronic decay channels $\DtoKpi$, $\DtoKpipi$, $\DstartoDpi\to\rm K^- \pi^+\pi^+$, and $\DstophipitoKKpi$. The Inner Tracking System (ITS) comprises six layers of silicon pixel, drift, and strip detectors, and is used for the tracking of charged particles and the reconstruction of the primary and secondary vertices. 
Track reconstruction in the central barrel ($|\eta| < 0.9$) starts from the large Time Projection Chamber (TPC), which also provides particle identification (PID) by measuring the specific energy loss $\dEdx$ of charged particles passing through it. Further PID information is given by the Time Of Flight detector (TOF) via the flight times of charged particles. The V0 scintillator array is used for the minimum-bias trigger. 
The Silicon Pixel Detector (SPD), the V0A detector, and the Zero-Degree Neutron Calorimeter (ZNA) are used to classify the charged-particle multiplicity and centrality of \pPb~collisions. 
D-meson candidates are selected based on PID of the decay tracks and the D-meson decay topology, for example by selecting on the impact parameters of the decay tracks and the pointing angle between the reconstructed D-meson momentum and its flight line.

The analysis of electrons from heavy-flavour hadron decays~\cite{ALICEhfeinclusivepPb} also uses the ITS, TPC and TOF. The Transition Radiation Detector (TRD) and the Electromagnetic Calorimeter (EMCal) provide further triggering and PID at intermediate to high $\pt$.
Background electrons from known hadronic decays are statistically subtracted from the inclusive electron yields, while photon conversions and Dalitz decays are removed via an invariant mass method. 

Muons from heavy-flavour hadron decays~\cite{ALICE_muons} are identified at forward and backward rapidity\footnote{Per ALICE convention, in \pPb collisions, the Pb-going direction is referred to as `backward' (negative) rapidity, and the p-going direction is referred to as `forward' (positive) rapidity.} using the muon spectrometer. This detector covers $-4.0 < \eta < -2.5$ and consists of a front absorber, a series of tracking chambers, a large dipole magnet, and a set of triggering chambers that sit behind a filter wall. As well as providing an effective trigger for muons, the trigger chambers also allow the rejection of background from punch-through hadrons. Muons from $\pi$ and K decays are subtracted via data-tuned Monte Carlo simulations.

\section{Results}
\label{sec:RpPb}

The nuclear modification factor, $\RpPb$, is calculated  as $\RpPb = \frac{\mathrm{d}\sigma_\mathrm{pPb}/\mathrm{d}\pt}{A\cdot\mathrm{d}\sigma_{\mathrm{pp}}/\mathrm{d}\pt}$,
where $A$ is the mass number of the Pb nucleus, and $\sigma_\mathrm{pPb}$ and $\sigma_\mathrm{pp}$ are, respectively, the particle production cross section in \pPb~collisions at $\sqrtsNN=5.02$~TeV and in pp collisions. Reference data were taken for  \pp collisions at $\sqrts=2.76$\,\TeV and $\sqrts=7$\,TeV, and scaled to $\sqrts=5.02$\,TeV using FONLL calculations~\cite{FONLL}.
\begin{figure}[h]
 \centering
	\includegraphics*[width=14.6pc,trim={0 0 1pc 0},clip]{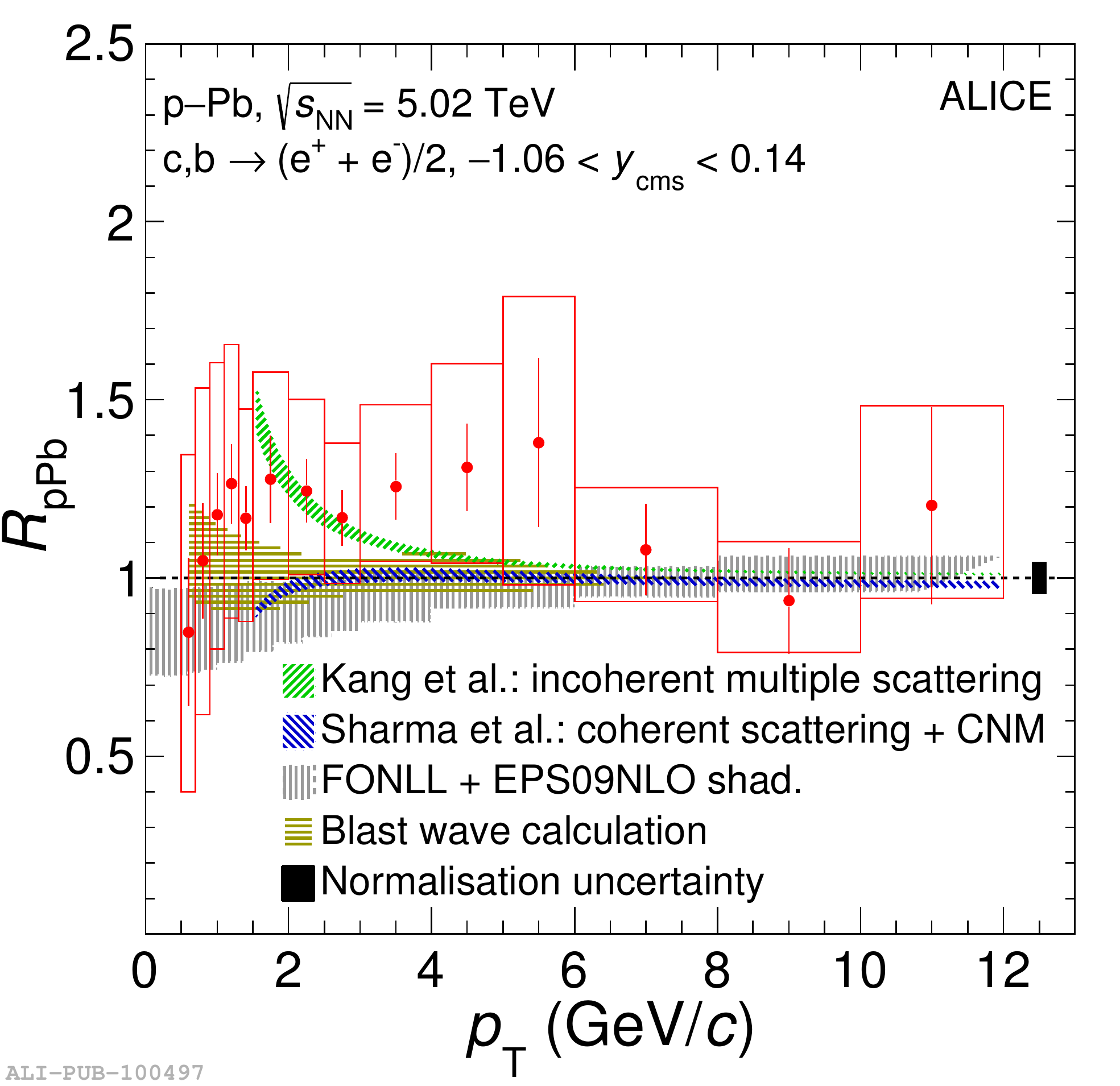} 
	\raisebox{0.4ex}{\includegraphics*[width=20pc,trim={0 0 0 1pc},clip]{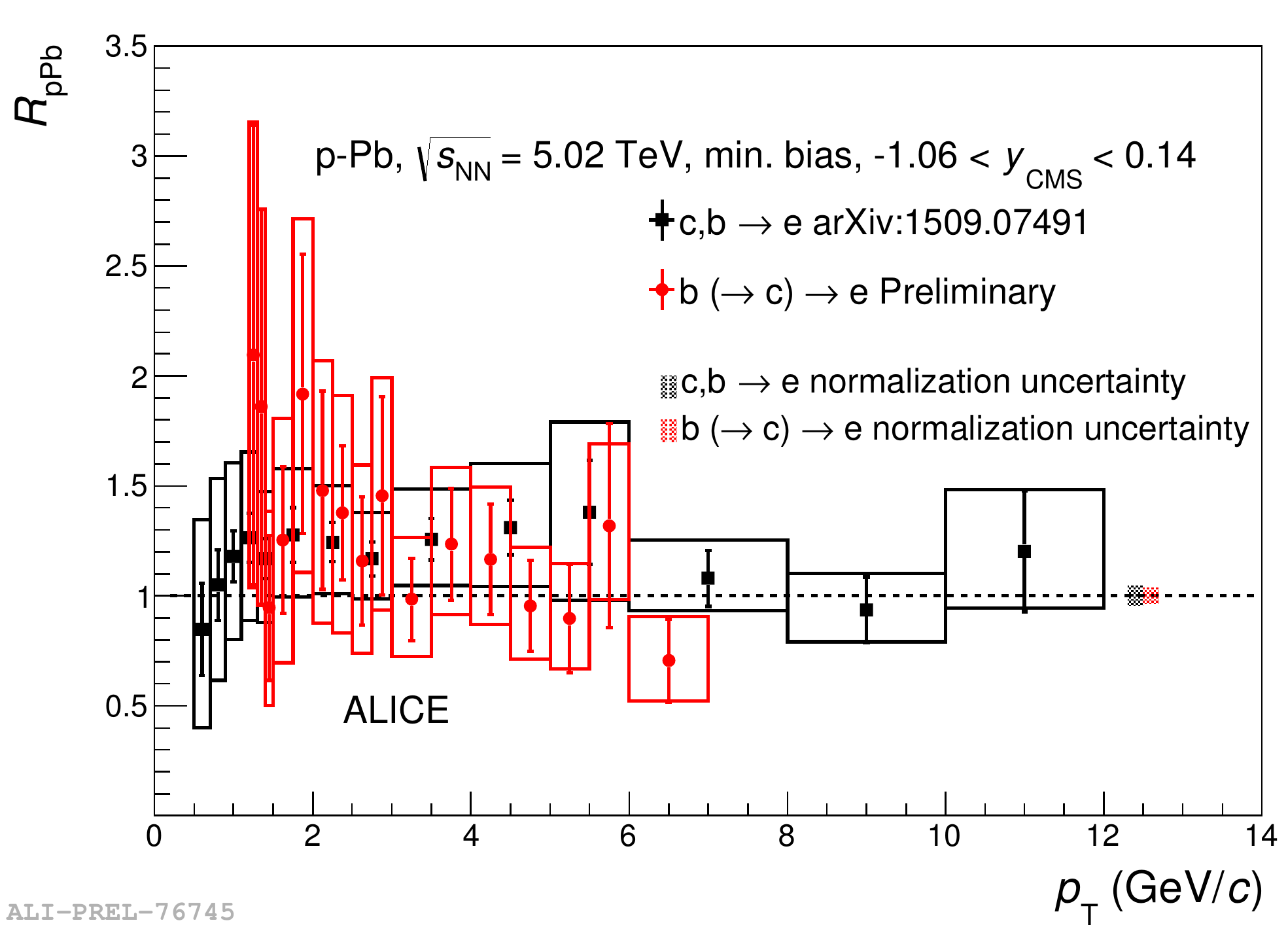}}
\caption{Left: \RpPb of inclusive electrons from heavy-flavour hadron decays~\cite{ALICEhfeinclusivepPb}, compared with models~\cite{FONLL,Sickles201451,Vitev,Kang201523,EPS09}. Right: Preliminary beauty-hadron decay electron \RpPb (red) compared with the overall \RpPb of electrons from heavy-flavour hadron decays (black).}
 \label{fig:electronRpPb}\label{fig:elec_RpPb_models}\label{fig:elec_RpPb_beauty}
\end{figure}

\Figref{fig:elec_RpPb_models} (left) shows the \RpPb of electrons from heavy-flavour hadron decays (i.e. those originating from both charm and beauty)~\cite{ALICEhfeinclusivepPb}.
The results are consistent with unity within uncertainties, implying that CNM effects have a negligible effect on heavy-flavour production at high \pt. The \RpPb is also reproduced by a variety of model calculations
~\cite{FONLL,Sickles201451,Kang201523,Vitev,EPS09}. 
It is also possible to separate the contributions of charm and beauty production via the impact parameter distributions of the decay products, as it is expected that this distribution will be broader for beauty than charm due to a larger separation between the primary and secondary vertices. The \RpPb of inclusive heavy-flavour and beauty-decay electrons are compared in~\figref{fig:elec_RpPb_beauty} (right); both results are consistent with unity and with one another within the uncertainties.

\begin{figure}[tb]
 \centering
	\includegraphics*[width=0.49\textwidth,trim={0 0 1.6cm 0},clip]{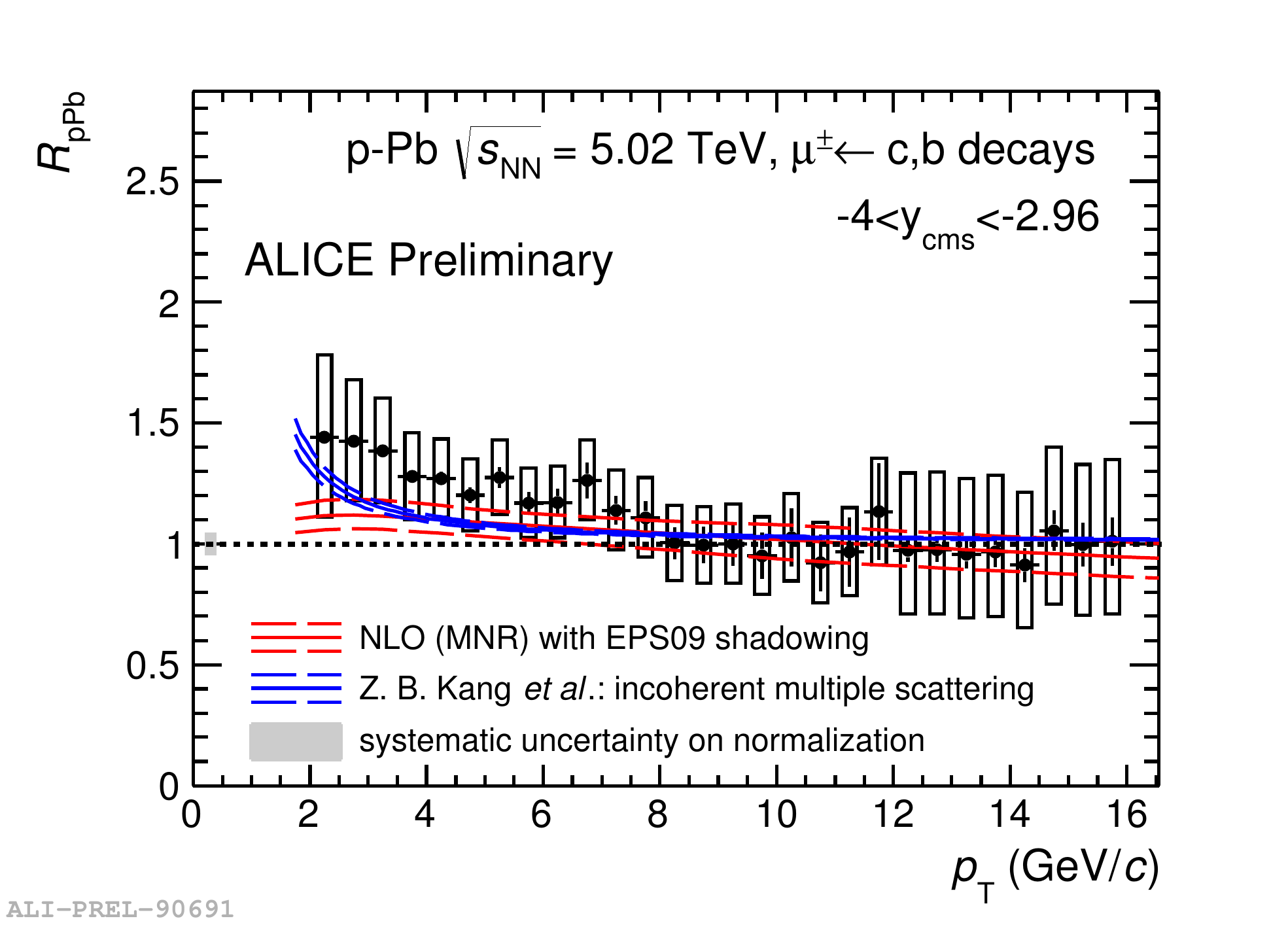}
	\includegraphics*[width=0.49\textwidth,trim={0 0 1.6cm 0},clip]{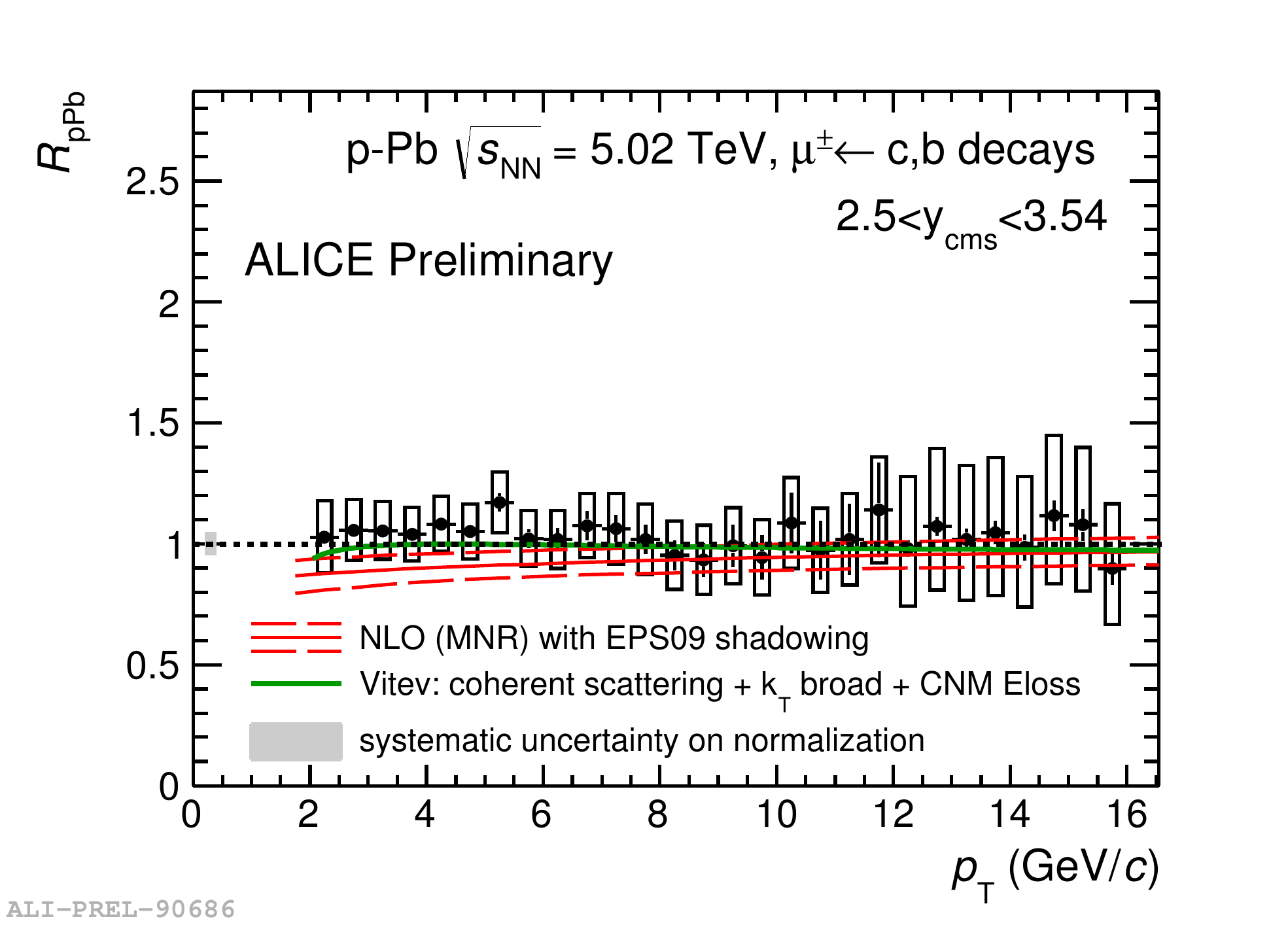}
\caption{\RpPb of muons from heavy-flavour hadron decays in two rapidity regions, compared with models~\cite{CNMEloss,MNR,EPS09,Kang201523}. Left: backward rapidity ($-4<y_\mathrm{cms}<-2.96$). Right: forward rapidity ($2.5<y_\mathrm{cms}<3.54$).}\label{fig:muon_RpPb_back}\label{fig:muon_RpPb_for}

\label{fig:muon_RpPb}
\end{figure}

The left (right) panel of \figref{fig:muon_RpPb} shows the \RpPb results for muons from heavy-flavour hadron decays at backward (forward) rapidity. The \RpPb results at forward rapidity are consistent with unity at all \pt, as was observed for electrons from heavy-flavour hadron decays at mid-rapidity. At backward rapidity, the \RpPb is also consistent with unity for $\pt > 4~\GeVc$, however there is an enhancement above unity for $2<\pT<4~\GeVc$. The results in both rapidity regions are replicated within uncertainties by models~\cite{CNMEloss,MNR,EPS09,Kang201523}.

\begin{figure}[h!]
 \centering
	\includegraphics[width=0.329\textwidth]{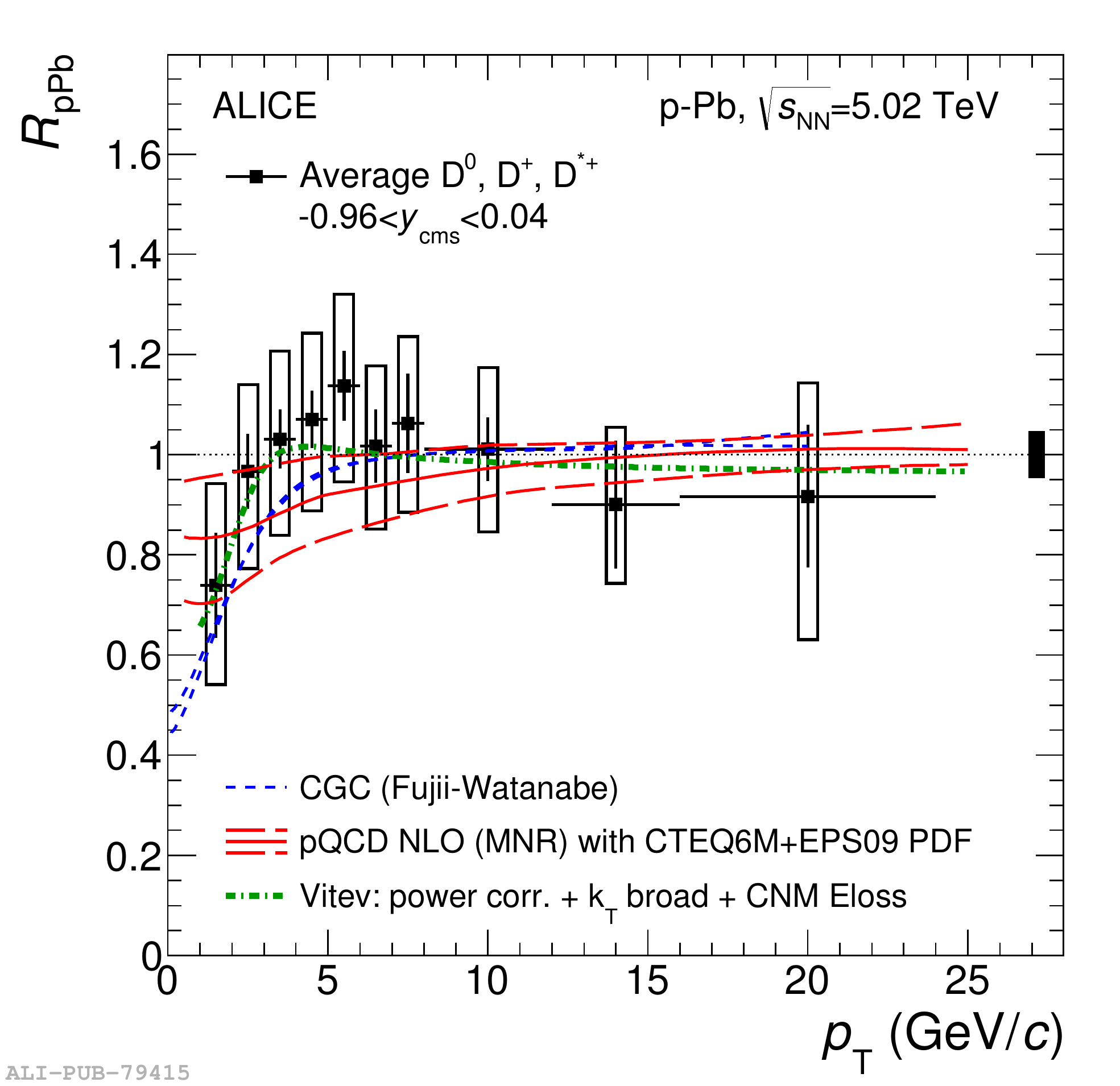}
	\includegraphics[width=0.329\textwidth]{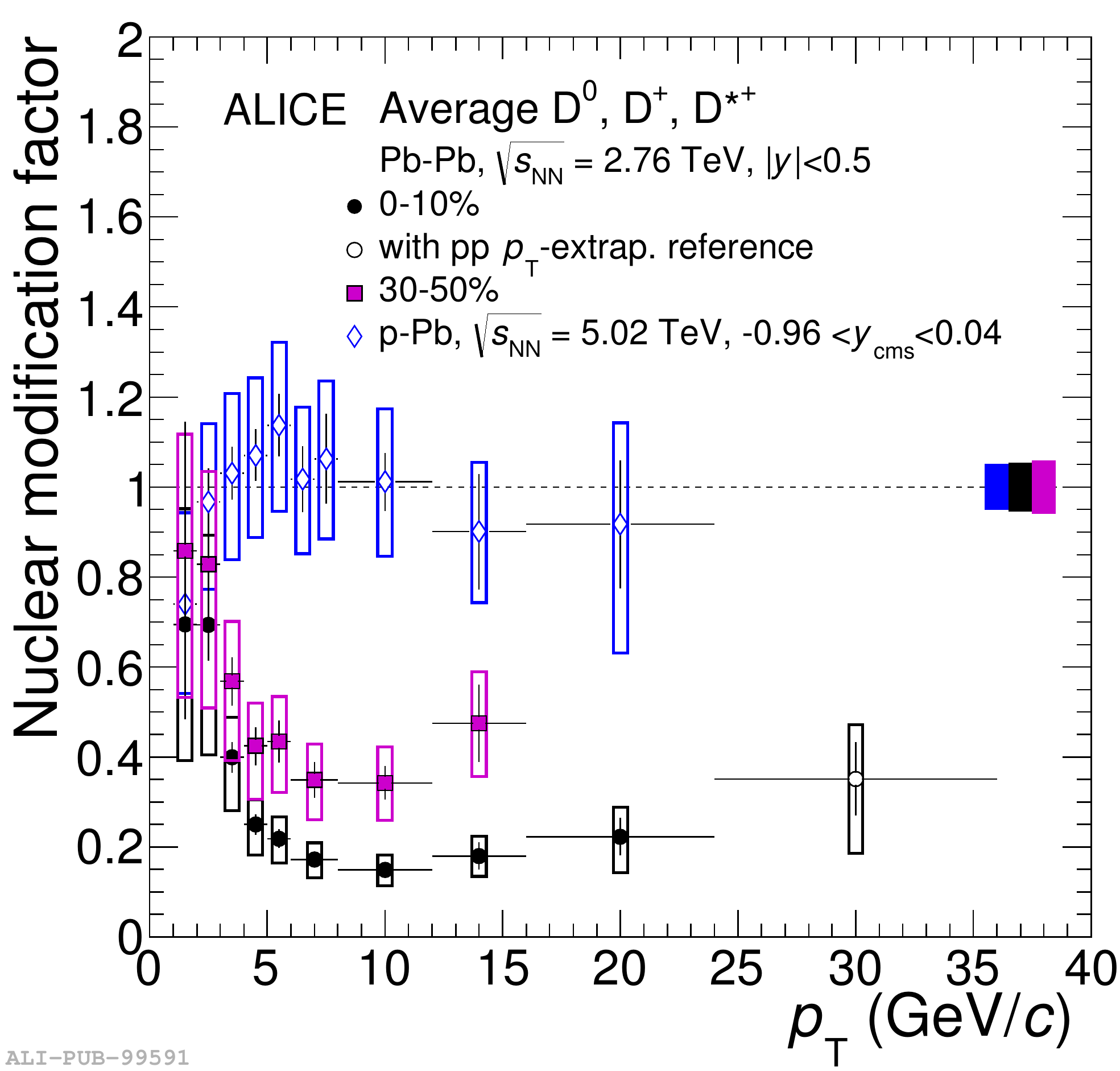}
	\includegraphics*[width=0.329\textwidth]{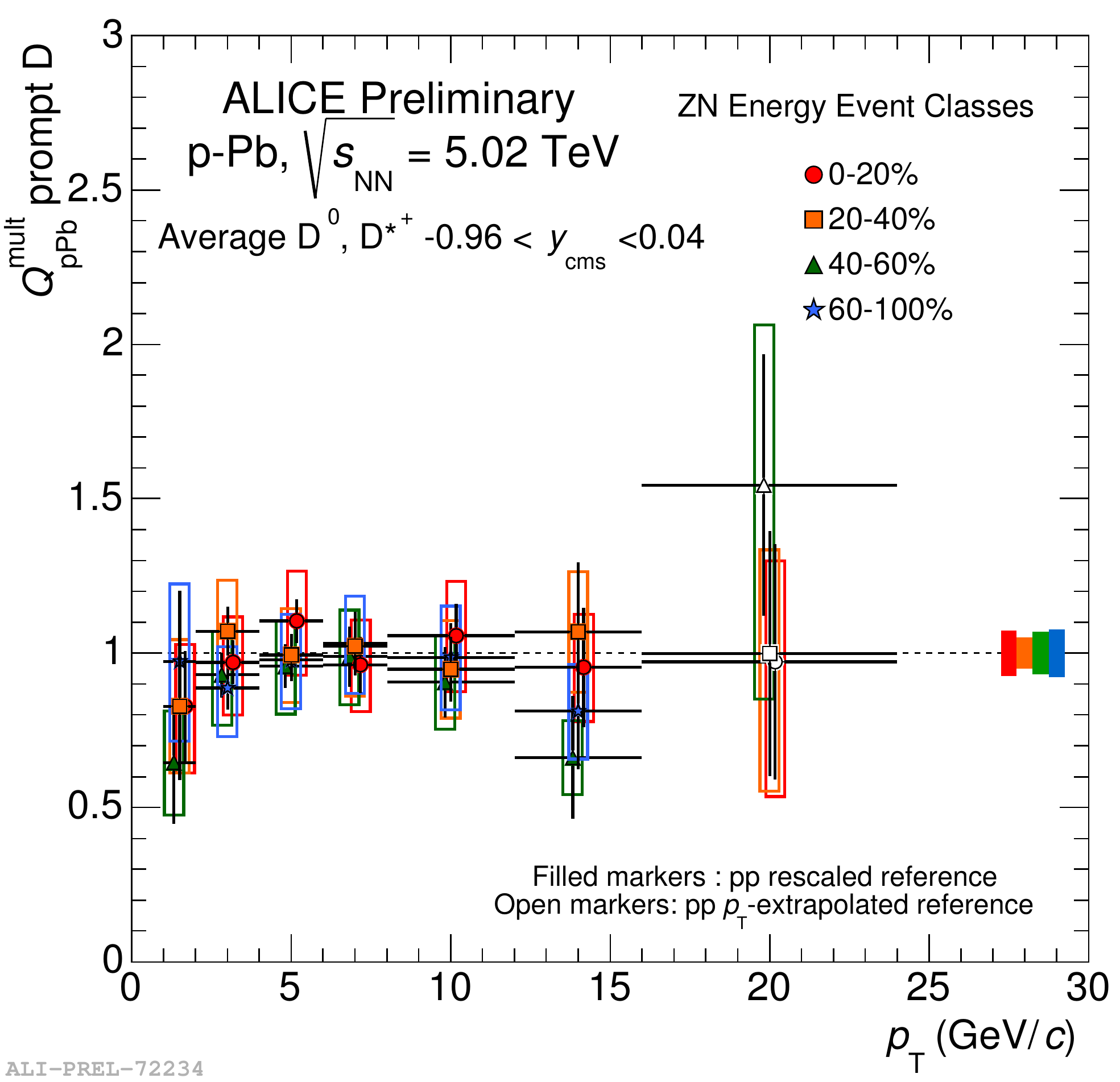}
	\caption{Average \RpPb of $\Dzero$, $\Dplus$ and $\Dstar$ mesons~\cite{ALICEDmespPbmb} compared with (left) models that include initial-state effects~\cite{MNR,EPS09,CTEQ6M,CGC,Vitev} and (middle) \RAA at $\sqrtsNN = 2.76~\TeV$~\cite{ALICEDmesonRAApT}. Right: average D-meson \QpPb results in four centrality classes.}\label{fig:Dmes_RpPb_RAA} \label{fig:Dmes_RpPb_models}
\label{fig:DmesonRpPb}\label{fig:Dmes_QpPb}
\end{figure}

\Figref{fig:Dmes_RpPb_models} (left) shows the average $\RpPb$ of prompt $\Dzero$, $\Dplus$ and $\Dstar$ mesons as a function of $\pt$
~\cite{ALICEDmespPbmb}. 
The \RpPb of D mesons is consistent with unity for $\pT > 2$~\GeVc, and is described by calculations that include CNM  effects~\cite{MNR,EPS09,CTEQ6M,CGC,Vitev}.
By contrast, the nuclear modification factor in \PbPb~collisions, \Raa (\figref{fig:Dmes_RpPb_RAA} (middle))~\cite{ALICEDmesonRAApT}, shows a significant suppression for $\pT > 3$~\GeVc in both the 0--10\% and 30--50\% centrality classes.
As the \RpPb remains consistent with unity over the entire measured \pt range, it is concluded that the suppression in \PbPb collisions occurs due to final-state effects in the medium rather than CNM effects.

Further studies of heavy-flavour production as a function of multiplicity and centrality allow the interplay of hard and soft processes to be examined. One approach is to measure the centrality-dependent nuclear modification factor, \QpPb. \QpPb is defined as $\QpPb^\mathrm{mult} = \frac{\mathrm{d}N^\mathrm{D}_\mathrm{pPb}/\mathrm{d}\pt}{\langle\TpPb^\mathrm{mult}\rangle\cdot\mathrm{d}\sigma^\mathrm{D}_\mathrm{pp}/\mathrm{d}\pt}$. $\TpPb$ denotes the nuclear overlap function, which is calculated under the assumption that the charged-particle multiplicity at mid-rapidity scales linearly with the number of participant nucleons~\cite{ALICEpPbcent}. The collision centrality is estimated using the ZNA calorimeter. The average D-meson \QpPb (\figref{fig:Dmes_QpPb} (right)) is consistent with unity, and is independent of both centrality and \pt within uncertainties.
Finally, we consider the relative D-meson yields as a function of charged-particle multiplicity. The relative yield is defined as the corrected yield per event in a given multiplicity class, divided by the multiplicity-integrated yield per event.
 Two multiplicity estimators were used: the number of track segments reconstructed in the SPD  at mid-rapidity ($|\eta| < 1.0$), and the V0A signal at backward rapidity ($2.8<\etalab<5.1$).
\begin{figure}[h!]
 \centering
	\includegraphics*[width=0.329\textwidth]{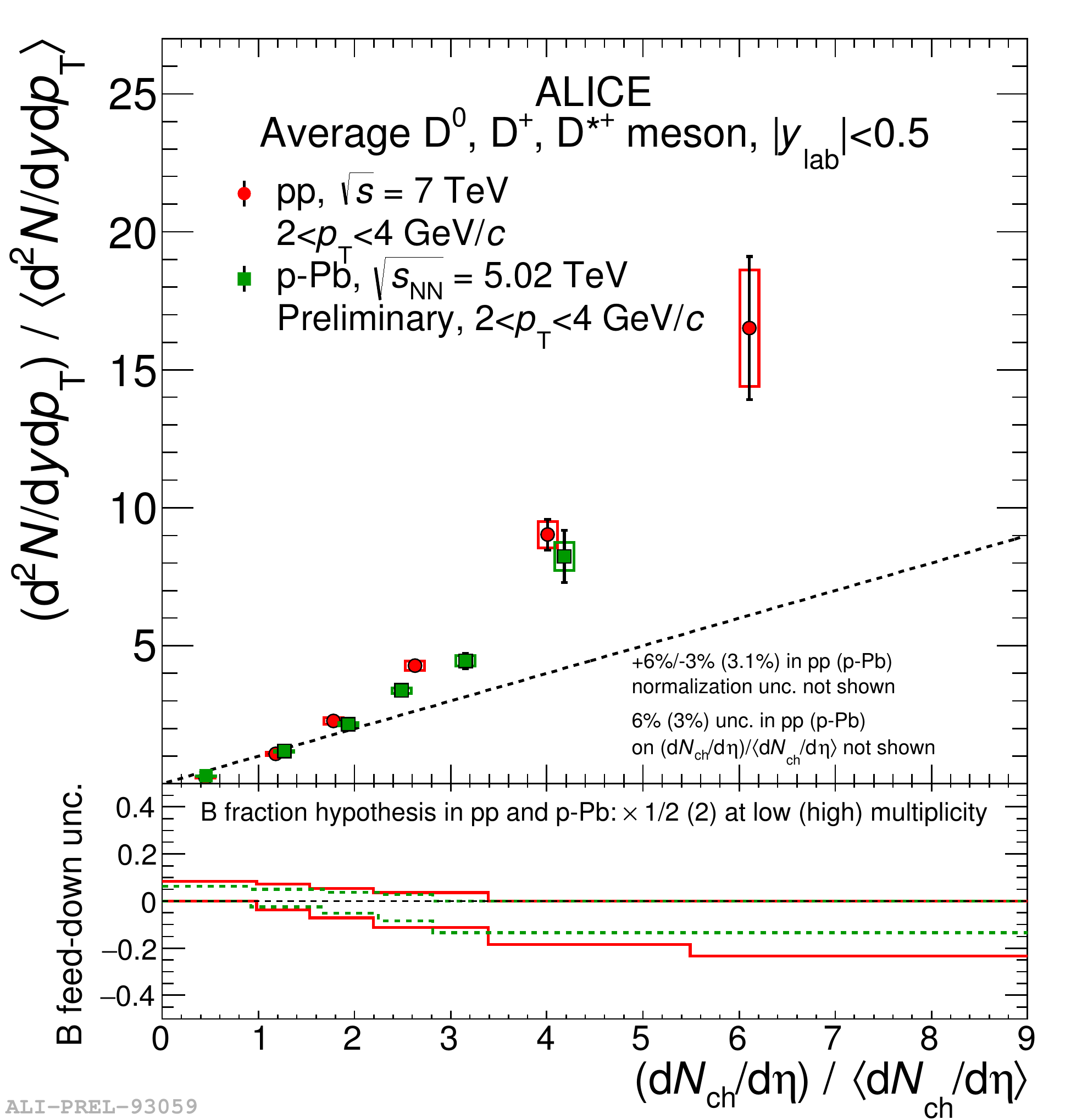}
	\includegraphics*[width=0.329\textwidth]{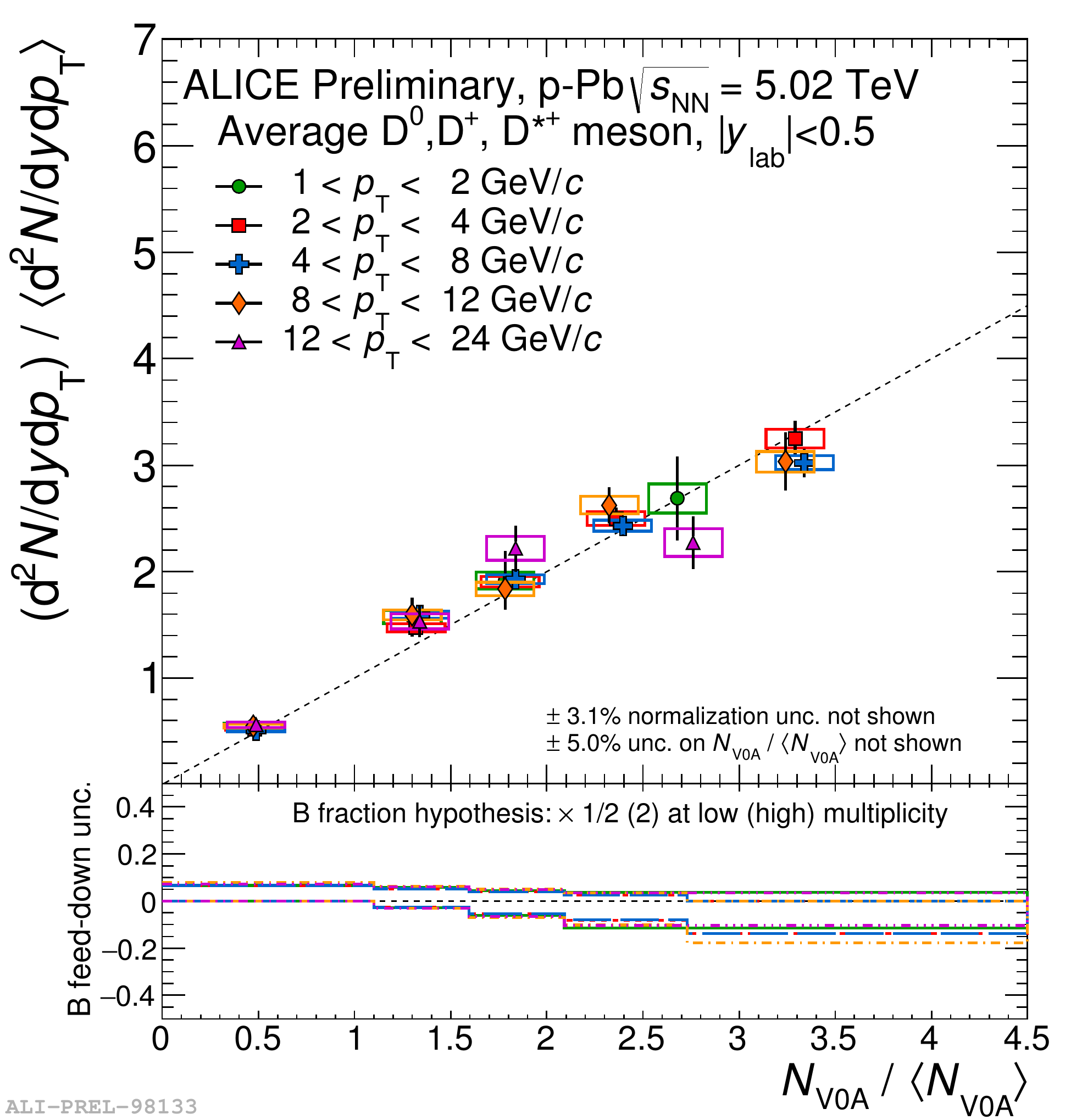}
	\raisebox{-0.5ex}{\includegraphics*[width=0.329\textwidth]{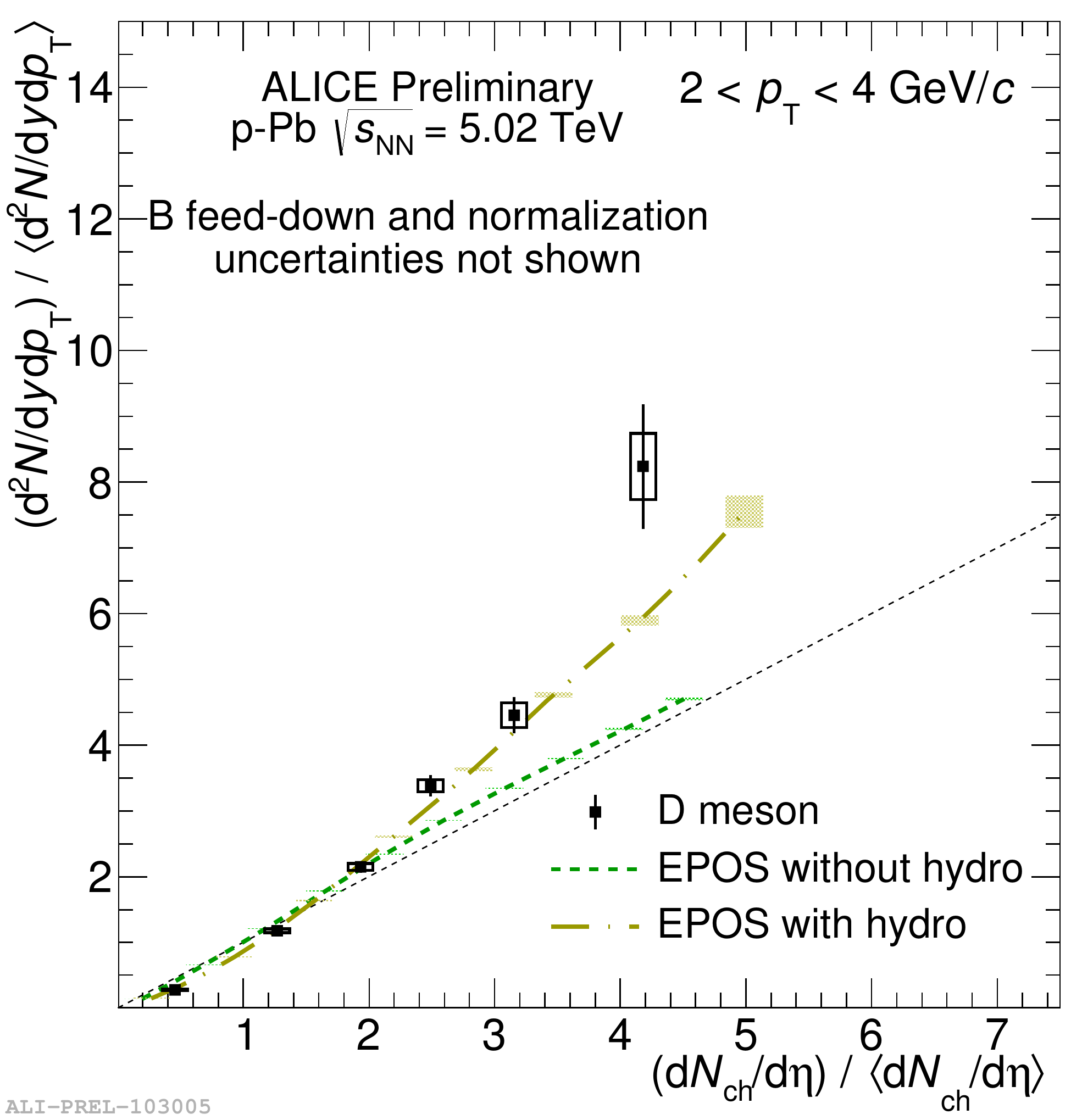}}
\caption{Relative D-meson yields as a function of multiplicity, estimated with the (left) SPD and (middle) V0A estimator. The pp results in the left-hand plot are taken from~\cite{ALICEDmesvsmultpp}. The dashed black lines represent a linear increase ($y=x$), and are displayed to guide the eye.
Right: comparison of multiplicity-dependent results from the SPD estimator with EPOS 3 calculations~\cite{EPOS3_1,EPOS3_2}, with and without viscous hydrodynamics.}
\label{fig:DvsMult_1}\label{fig:Dmes_mult_vza}\label{fig:Dmes_mult_mid_pp}\label{fig:Dmes_mult_model}
\end{figure}
\Figref{fig:Dmes_mult_mid_pp} (left) shows the average relative yields obtained for \Dzero, \Dplus and \Dstar mesons with the SPD estimator for $2<\pt<4~\GeVc$, compared with the equivalent measurement in pp collisions~\cite{ALICEDmesvsmultpp};~\figref{fig:Dmes_mult_vza} (middle) shows the results from the V0A estimator in five \pt bins. For the SPD estimator, the results in both \pp and \pPb collisions exhibit a similar multiplicity dependence, with each showing a faster-than-linear relative D-meson yield increase 
at high multiplicity. The faster-than-linear increase seen in \pp collisions is attributed to multi-parton interactions~\cite{ALICEDmesvsmultpp}. The results from the V0A estimator in \pPb collisions, on the other hand, show a roughly linear increase as a function of multiplicity. 
The results from the SPD estimator in \pPb collisions are also compared with EPOS 3 calculations~\cite{EPOS3_1,EPOS3_2} in~\figref{fig:Dmes_mult_model} (right). It was found that the multiplicity dependence of D-meson production is described better by calculations that include viscous hydrodynamics, especially at higher multiplicity.

%% References

\bibliographystyle{elsarticle-num}
\bibliography{proceedings}

\end{document}